%% file: main.tex
\documentclass[10pt]{llncs}

    \def\showcomments{0}

    \input{macros.sty}

    \title{A Protocol for Cast-as-Intended Verifiability with a Second Device}

    \author{
        Johannes M\"uller \inst{1}
        \and
        Tomasz Truderung \inst{2}
    }

    \institute{
        University of Luxembourg
        \and
        Polyas GmbH, Germany
    }

\begin{document} 

    \maketitle

\input{introduction.tex}

\input{related-work.tex}

    \input{generic-protocol.tex}

\input{security.tex}

    \input{instantiations.tex}

    \appendix
    
    \input{zkp.tex}

    \input{zkp-eqlog.tex}

    \bibliography{main}
\end{document}

%% file: introduction.tex
\begin{abstract}
    Numerous institutions, such as companies, universities, or
    non-govern\-mental organizations, employ Internet voting for remote
    elections. Since the main purpose of an election is to determine the voters'
    will, it is fundamentally important to ensure that the final election result
    correctly reflects the voters' votes. To this end, modern secure Internet
    voting schemes aim for what is called \emph{end-to-end verifiability}. This
    fundamental security property ensures that the correctness of the final
    result can be verified, even if some of the computers or parties involved
    are malfunctioning or corrupted.

    A standard component in this approach is so called \emph{cast-as-intended
    verifiability} which enables individual voters to verify that the ballots
    cast on their behalf contain their intended choices. Numerous approaches for
    cast-as-intended verifiability have been proposed in the literature,
    some of which have also been employed in real-life Internet elections.
    These different approaches strike a balance between practical
    aspects and security guarantees in different ways.

    One of the well established approaches for cast-as-intended verifiability
    is to employ a second device which can be used by voters to audit their
    submitted ballots. This approach offers several advantages---including support for flexible ballot/election types and
    intuitive user experience---and it has been used in  real-life elections, for
    instance in Estonia~\cite{heiberg2016improving}. Importantly, the solutions
    based on this approach are typically not bound to a particular election
    protocol, but rather they can augment many existing and practically relevant
    voting protocols.

    In this work, we improve the existing solutions for cast-as-intended
    verifiability based on the use of a second device. We propose a solution
    which, while preserving the advantageous practical properties sketched above,
    provides tighter security guarantees. Our method does not increase the risk
    of vote-selling when compared to the underlying
    voting protocol being augmented and, to achieve this, it requires only
    comparatively weak trust assumptions. It can be combined with various voting
    protocols, including commitment-based systems offering everlasting privacy.

    In summary, our work presents a new option to strengthen cast-as-intended
    and thus end-to-end verifiability of real-world Internet elections.
\end{abstract}

\section{Introduction}

Interned voting has been employed by numerous institutions, such as companies,
universities, or non-govern\-mental organizations, as well as for some remote
elections on the national level. The adoption of Internet voting has been driven by
several practical benefits of this form of voting, in particular enabling all voters
to participate regardless of their physical location.

Internet voting comes, however, with its own challenges and risks. One of those risks lies in potential malfunctioning, which for such complex software-hardware systems cannot be easily ruled out. Such issues can be caused by design/programming mistakes, security vulnerabilities, or even by deliberate tampering with the deployed system. In any case, malfunctioning can
potentially have serious practical consequences. In fact, if the final election result is accepted although it does not correspond to the votes submitted by the voters, then the actual purpose of an election is undermined.

To safeguard against such risks, modern Internet voting systems strive for so-called \emph{end-to-end verifiability}~\cite{cortier2016sok}. This fundamental property requires the system to provide \emph{evidence} that
the election result accurately reflects the votes cast by eligible voters. Importantly, such evidence must be independently
verifiable.

\emph{Individual verifiability} is an essential part of end-to-end verifiability. This property guarantees that each individual voter is able to verify whether the vote she entered to her voting device is in fact tallied. Individual verifiability is typically achieved as follows. First, the voter verifies whether her (possibly malfunctioning) voting device cast her encrypted vote as she intended; this feature is called \emph{cast-as-intended verifiability}. Then, the voter checks whether the ballot she cast is tallied by the authorities; this feature is called \emph{tallied-as-recorded verifiability}. If both of these features are in place, they enable all individual voters to verify independently that exactly their secret votes are tallied.

The requirement of end-to-end verifiability in general, and
individual verifiability in particular, is not
only widely stipulated by the research community, but it is also
becoming part of the standard legal requirements and frameworks.
The relevance of verifiability is
recognized, for instance, by the Council of Europe in its
recommendation on standards for e-voting \cite{CoE-2017}.
Importantly, the same document specifies that ``individual verifiability can be implemented provided adequate safeguards exist to prevent coercion or vote-buying''. Requirements for individual verifiability are also
postulated for the Swiss elections in \emph{Federal Chancellery
Ordinance on Electronic Voting}\footnote{
    ``For the purpose of individual verification,
voters must receive proof that the server system has
registered the vote as it was entered by the voter on the
user platform as being in conformity with the system.''
\cite{VEleS-2013}
},
for the Estonian elections in \emph{Riigikogu Election Act}\footnote{``A voter has an opportunity to verify whether the application used for electronic voting has transferred the vote cast by the voter to the electronic voting system according to the voter's wish.'' \cite{REA-2002}
},
and for non-political elections in Germany
\cite{BSI-TR-03162}.

Numerous techniques for cast-as-intended verifiability have been proposed in the
literature (see,
e.g.,~\cite{GalindoGP15,Haenni16,HaenniKLD17,BrelleTruderung2017cast,cortier2019beleniosvs,benaloh2006simple,heiberg2016improving,guasch2016challenge,grewal2015vote,kusters2016select,ryan2016selene}).
Some of them are also employed in real elections, for
example~\cite{heiberg2016improving} in the Estonian voting system IVXV,
and~\cite{benaloh2006simple} in the Helios voting
system~\cite{Adida-Security-2008}. Each of these techniques provides its own
balance between security, trust assumptions, usability, and deployability.


\subsection{Our contributions}

We propose a method for cast-as-intended verifiability that offers a new
balance between security guarantees and practical aspects; in particular, it can be used to
augment many relevant Internet voting protocols. Our method does not increase
the risk of vote selling, when compared to the underlying voting protocol
being augmented, which is provided under comparatively weak trust assumptions.

More specifically, we optimized our cast-as-intended mechanism for the following
\emph{design goals}. The first four design goals \dg1-\dg4 are functional and
they essentially determine the election scenarios in which the cast-as-intended
mechanism can be applied; in combination, the functional design goals cover a
wide range of real-world elections over the Internet, which is a central
requirement for our practically orientated work. The last two design goals
\dg5-\dg6 express security features that the cast-as-intended mechanism
should provide.

\begin{itemize}
    \def\dgb#1{\emph{\dg#1}}
    \item \dgb1 \emph{Support for flexible ballot types.} 
        The mechanism should not be restricted to only some
        specific ballot types, such as simple ballots with
        relatively small number of candidates or simple ballot
        rules. On the contrary, it is desirable that complex
        ballots are supported, including, for instance, ballots
        with write-in candidates or ranked voting.

    \item \dgb2 \emph{Low cost}. The mechanism should not
        significantly increase the cost of the election, for
        instance by requiring dedicated secure printing/distribution 
        facilities.

    \item \dgb3 \emph{No disenfranchisement of voters}.
        The mechanism should not make unrealistic assumptions
        about voters' knowledge, abilities, and what they
        possess.  This rules out mechanisms which assume some
        sort of custom hardware. Also, the mechanism should be
        reasonably intuitive so that an average voter could
        understand what he/she is supposed to do and why.

    \item \dgb4 \emph{Modularity.} The mechanism can be used augment a large class of Internet voting protocols, in particular protocols using different type of tallying, and protocols with everlasting privacy. The method should support modular security analysis, where the security properties of the combined scheme can be derived from the security properties of the underlying protocol (without individual verifiability) and the properties of the individual verifiability method.

    \item \dgb5 \emph{No facilitation of vote-selling}. The mechanism should
    not make vote-selling easier than in the voting scheme being augmented.
    To be clear: We do not aim at protecting the overall voting scheme
    against vote-selling, but we require that the cast-as-intended mechanism
    should not additionally provide voters with receipts that they can use to
    \emph{trivially} prove towards a vote-buyer how they voted.

    \item \dgb6 \emph{Possibly minimal trust assumptions.} 
    We prioritize solutions which require weaker or more flexible trust
    assumptions.

    An example of such a trust assumption is reliance on some trapdoor values
    generated by a trusted entity, where for the integrity of the individual
    verifiability method, we need to assume that this party is honest (not
    corrupted) and that the trapdoor value does not leak.

 \end{itemize}

As we discuss in detail in Sec.~\ref{sec:related-work}, no existing method
for cast-as-intended verifiability in the literature achieves all of our
design goals simultaneously in a satisfactory degree. We note, however, that,
while our solution is optimized for our particular design goals, other
methods may be better suited for different election settings which require
different resolution of the security/usability/deployability trade-offs.


\medskip\noindent
Let us now explain on a high level how and why our cast-as-intended mechanism achieves all of our design goals satisfactorily:

\begin{itemize}
	\item We follow the approach that employs a second device, called \emph{audit device}, which voters can use to verify that the ballot submitted on their behalf contains their intended choice. This approach is established and it has already been used in real-life elections, for example in Estonia \cite{heiberg2016improving}. More precisely, in our method the voter can use a general-purpose device, such as a mobile phone or tablet, as the audit device. This audit device needs to be able to scan QR-codes and it also has to connect with the Internet in order to communicate with the election system. In this way, we avoid a costly additional infrastructure \dg2, and we do not need to make unrealistic assumptions about what voters possess \dg3.

	\item The audit procedure is straightforward from a voter's point of view, as explained next. Once the encrypted ballot has been sent to the election system, a QR-code is displayed by the voting application. The voter uses the audit device to scan this QR-code. The audit device then prompts the voter to authenticate against the election system and, if this authentication is successful, it shows the voter's choice in plaintext, in the same form as the ballot was displayed in the primary (voting) device. We note that, nowadays, most voters are used to such or similar checks, for example in the context of secure online banking. Furthermore, the audit step is optional and thus not required to cast a ballot successfully. In summary, we make reasonable assumptions about the voters' knowledge and abilities \dg2.

	\item On a technical note, our method works well with all possible ballot types, even very complex ones, satisfying \dg1. Moreover, our modular method can be used to augment a large class of relevant Internet voting protocols \dg4, and the computational cost of the ballot audit computations is very reasonable \dg2.

	\item Unlike all previous cast-as-intended mechanisms that employ a second device~\cite{heiberg2016improving,guasch2016challenge}, our method simultaneously satisfies \dg5 and \dg6. We achieve this by providing \emph{cryptographic denialability}, without introducing additional trust assumptions. To this end, we employ \emph{interactive} zero-knowledge proofs where any party, by definition, can easily simulate the protocol transcript without the knowledge of the plaintext or the encryption coin. We use well-understood and relatively simple cryptography: our method, in its essence, relies on the interactive zero-knowledge proof of correct re-encryption. This results in simpler security proofs, which is an additional important factor in building trust.
\end{itemize}

Technically, the main challenge that we needed to resolve was induced by the general limitations of QR-codes. As described above, the QR-codes in our method are used as the only communication channel between the voting application and the
audit device. However, QR-codes provide only very restricted communication capacity since they are one-way and of very limited bandwidth. Now, in order to implement an \emph{interactive} zero-knowledge proof in this restricted setting, we split the role of the prover between the voting application and the election system in such a way that the election system does not learn anything during this process, while doing most of the `heavy-lifting'. The role of the verifier, as usual for such
schemes, is played by the audit device.

We note that, since the audit device displays the voter's choice, it needs to be trusted for ballot privacy. This is also the case for all other techniques that employ a second device~\cite{heiberg2016improving,guasch2016challenge}. In general, cast-as-intended methods based on return or voting codes do not have this disadvantage; however, they fall short on other design goals (see Sec.~\ref{sec:related-work} for more details).

%
%
%
%
%
%


\subsection{Structure of the paper}
In the next section, we provide more details on the existing approaches for cast-as-intended verifiability.
We describe our cast-as-intended mechanism in Sec.~\ref{sec:generic-protocol} and we analyze
its security in Sec.~\ref{sec:security}. In Sec.~\ref{sec:individual-verifiability}, we embed our
cast-as-intended protocol in an example protocol which provides
full individual verifiability and state higher-level security
properties of this protocol.
Finally, in Sec.~\ref{sec:instantiations}, we discuss some
practical cryptographic instantiations of our approach.

For completeness, we recall the relevant definitions of the zero-knowledge
proofs in Appendix~\ref{sec:zkp} and provide a concrete instantiation
of an interactive zero-knowledge proof for equality of discrete logarithms in
Appendix~\ref{sec:zkp-eq}.

%% file: related-work.tex
\section{Related Work}\label{sec:related-work}

    Various mechanisms for individual (cast-as-intended)
    verifiability have been proposed in the literature, striking
    different balances between security, usability, and several
    other practical aspects of the ballot casting process. In this
    section, we provide a brief overview of such mechanisms and explain why none of them offers our desired security features \dg5-\dg6 in those real-world elections that we are interested in, as determined by \dg1-\dg4. In particular, we
    focus here only on methods used for \emph{Internet e-voting}
    (as opposed to on-site voting).

%

%
%
%
%
%
%
%


\paragraph{Return Codes.}

    In the return-codes-based approach (see, e.g.,
    \cite{GalindoGP15,Haenni16,HaenniKLD17,BrelleTruderung2017cast}),
    before the voting phase starts, each voter receives a
    \emph{code sheet} (e.g. via postal mail) listing all the
    possible voting choices together with corresponding
    \emph{verification codes}. These codes are unique for each
    voter and should be kept secret.  During the voting phase,
    the voter, after having cast her ballot, receives (via the
    voting application or another dedicated channel) the return
    code corresponding to the selected choice. The voter compares
    this code to the one listed on the code sheet next to the
    intended choice.

    While this approach may work well and seems intuitive from the
    voter's point of view, it has several drawbacks. It does
    not scale well to complex ballots \nodg1, such as ballots with many
    candidates or when voters have the option to select multiple
    choices, because the code sheets become very big and the user
    experience quickly degrades (see, e.g.,~\cite{DBLP:conf/fc/KulykVMR20}). Another disadvantage is the
    cost incurred by (secure) printing and delivery of code
    sheets \nodg2. Finally, the printing and delivery facilities must
    be trusted in this approach: if the verification codes leak
    to the adversary, the integrity of the process completely
    breaks (a dishonest voting client can cast a modified choice
    and return the code corresponding to the voter's intended
    choices). This trust assumption is rather strong \nodg6.

\paragraph{Voting Codes.}

    In this approach, the voter, as above, obtains a voting sheet
    with voting codes. The difference is that the codes are not
    used to check the ballot after it has been cast, but instead
    to prepare/encode the ballot in the first place: in order to
    vote, the voter enters the code (or scans a QR-code)
    corresponding to their choice. By construction, the voting
    client is then only able to prepare a valid ballot for the
    selected choice and no other ones.  This approach is used,
    for example, in \cite{cortier2019beleniosvs}, where the
    voting codes are used not only to provide individual
    verifiability, but also to protect ballot privacy against
    dishonest voting client.


    This approach, similarly to the return codes, works only for
    simple ballots \nodg1; arguably, the usability issues are even
    bigger than for return codes, as the voter needs to type
    appropriated codes or scan appropriate QR-codes in order to
    correctly cast a ballot, not just compare the returned code
    with the expected one. As before, it incurs additional costs
    \nodg2 and requires one to trust the printing/delivery
    facilities \nodg6. 

\paragraph{Cast-\emph{or}-Audit.}

    The cast-or-audit approach, used for instance in Helios
    \cite{Adida-Security-2008}, utilizes the so-called Benaloh
    Challenge \cite{benaloh2006simple} method.  In this approach,
    the voter, after her choice has been encrypted by
    the voting client, has two options: she can \emph{either} choose to
    (1) cast this encrypted ballot \emph{or} (2) challenge (i.e., audit) the
    ballot.  If the latter option is chosen, the voting client
    enables the audit by revealing the randomness used to encrypt
    the ballot, so that the voter (typically using some
    additional device or application) can check that it contains
    the intended choice. The voter then starts the ballot cast
    process over again, possibly selecting a different choice.

    The security of this approach relies on the assumption that
    the client application (the adversary) does not know
    beforehand, whether the encrypted vote will be audited or
    cast.  Therefore, if the adversary tries to manipulate the
    ballot, it risks that this will be detected.  Note, however,
    that the ballot which is actually cast is not the one which
    is audited. This, unlike most of the other approaches,
    provides the voter with some (probabilistic) assurance, but
    not with fully effective guarantees.

    This method has the advantage that it does not produce a
    receipt (the voter can choose different candidates for the
    audited ballots) and that the audit device does need to have Internet access for verification (unlike cast-and-audit methods like ours), but it has several usability issues. The
    studies on usability of this
    scheme~\cite{weber2009usability,karayumak2011user} conclude
    that voters tend to not verify their votes and have serious problems
    with understanding the idea of this type of ballot audit,
    which make this approach score low on \dg3.
    The above issues render the cast-or-audit approach
    ineffective in practice.

\paragraph{Cast-\emph{and}-Audit.}

    The solution presented in this paper belongs to this
    category.  In this approach, the voter audits, typically
    using a second device, the cast ballot (before or after it is
    cast).

    This approach is used by the system deployed for the Estonian
    elections \cite{heiberg2016improving}. In this case, the
    voters can use a mobile application to scan a QR-code displayed
    by the voting client application. This QR-code includes the 
    random encryption coin used to encrypt the voter's choice.
    The audit device fetches the voter's ballot from the ballot
    box and uses the provided randomness to extract the voter's
    choice which is then displayed for the voter to inspect.

    This method is flexible as it works well also for complex
    ballot types \dg1 (the audit device conveniently displays the
    vote in the same way the ballot appeared in the main
    voting device).  The user experience, for this method, 
    is relatively simple \dg3. The method does not incur extra cost \dg2.

    The main disadvantage of this method in general is that the additional
    (audit) device must be trusted for ballot privacy, as it
    ``sees'' the voter choice in clear. Also, the fact that the
    voters need to have an additional device (such as a mobile
    phone), which is able to scan QR-codes and
    which has Internet access, can be seen as a disadvantage.
    However, with the high availability of such devices, this
    does not seem to be a significant issue in practice.  
    The correctness of the ballot audit process relies on the
    assumption that one of the devices the voter uses (either the
    main voting device or the audit device) is not corrupted. In
    practice, it is therefore desirable that the software
    programs (apps) run on these two devices were developed and
    installed independently, ideally by different vendors or
    trusted third parties (e.g., pro-democratic organizations).

    The main idea of the cast-as-intended mechanism proposed
    in~\cite{heiberg2016improving} is that the QR-code includes the encryption
    random coins.
    Such a coin constitutes a trivial and direct evidence for
    the plaintext content of the encrypted ballot. As such, the simple
    cast-as-intended mechanism of~\cite{heiberg2016improving} does not provide
    cryptographic \emph{deniability} and may potentially facilitate vote
    buying/coercion \nodg4. Whether this potential for vote buying/coercion
    becomes an actual threat depends on the overall voting protocol; for
    instance, the Estonian system allows for vote updating as a measure to
    mitigate the threat of coercion. The lack of cryptographic deniability
    remains nevertheless a serious drawback of this method and significantly
    limits it applicability.

    The issue of selling cast-as-intended data as trivial receipts in Internet
    elections is addressed in
    \cite{guasch2016challenge}, where cryptographic deniability
    is provided using non-interactive zero-knowledge proofs with
    trapdoors.  This solution to the receipt problem has,
    however, its own issues: the trapdoor (for each voter) is
    generated by a registrar who therefore needs to be trusted
    for integrity of this method. This is arguably a strong
    trust assumption \nodg6.

    As already mentioned in the introduction, the solution
    presented in this paper, while also providing cryptographic
    denialability, does not require such an additional trust
    assumption \dg6. It also avoids the relatively complex
    cryptographic machinery of \cite{guasch2016challenge}, which
    often is the source of serious programming flaws (see,
    e.g.,~\cite{DBLP:conf/sp/HainesLPT20}).


\paragraph{Custom hardware tokens.}

    Some other solutions, such as \cite{grewal2015vote}, rely on
    using dedicated hardware tokens during the cast process.
    Relying on custom hardware makes these solutions expensive
    and difficult to deploy in real, big scale elections \nodg2,
    \nodg3.  Furthermore,~\cite{DBLP:conf/eurosp/KremerR16}
    demonstrated that~\cite{grewal2015vote} suffers from several
    security issues and concluded that~\cite{grewal2015vote} was
    not yet ready to be deployed.

\paragraph{Tracking codes.}







    The sElect system \cite{kusters2016select} achieves
    cast-as-intended in a simple way: voters are given random
    tracking numbers as they cast their ballots. After the tally, voters can check
    that their tracking numbers appear next to their respective votes.

    This method is simple and intuitive for the voters, but has
    the following drawbacks. End-to-end verifiability relies on the voters to
    perform the checks because there is no universal verifiability
    process that complements the individual verifiability made by
    the voters. Also, in \cite{kusters2016select}, the 
    tracking codes were ``generated" and entered by the voters.
    This is somehow problematic both from the usability point of
    view and because of the poor quality of ``random" numbers
    made up by voters (see, e.g.,~\cite{DBLP:phd/ethos/Bonneau12}). 
    Altogether, this method seems to take somehow unrealistic
    assumptions about the voters: that the voters carry out the
    process often enough for achieving the desired security level
    and that they are able to generate decent randomness \nodg3.
    Furthermore, the tracking codes, as used in \cite{kusters2016select}, may
    allow for simple vote buying \nodg5.

    The construction presented in Selene~\cite{ryan2016selene} also
    builds upon the idea of tracking codes, but further
    guarantees receipt-freeness, and thus impedes vote buying,
    due to a complex cryptographic machinery. The
    cast-as-intended mechanism is here, however, tightly bound to
    the e-voting protocol and thus not modular \nodg4.
    In particular, unlike for the method proposed in
    this paper, it is not immediately obvious how to improve
    Selene towards everlasting privacy or how to instantiate
    Selene with practical post-quantum primitives.

%% file: generic-protocol.tex
\section{Cast-As-Intended Verifiability: Generic Protocol}\label{sec:generic-protocol}

In this section, we present our protocol for cast-as-intended
verifiability. We take a modular approach: in Section \ref{subsec:generic-protocol-basic}, we start off
describing a generic basic ballot submission process \emph{without}
cast-as-intended, and then, in Section
\ref{sec:generic-protocol-extended}, we build upon this basic
process and extend it with cast-as-intended verifiability.

In Sec.~\ref{sec:individual-verifiability}, we will explain how to (easily) extend our cast-as-intended protocol so that full individual verifiability is achieved.

\subsection{Basic ballot submission}\label{subsec:generic-protocol-basic}

We describe now how the basic ballot submission (sub-)protocol of
an e-voting protocol without cast-as-intended verifiability works
which establishes the starting point for our mechanism,
introduced in the next subsection.  Doing this, we abstract from
some aspects (such as authentication) which are irrelevant for
our cast-as-intended protocol.

We provide this explicitly defined basic protocol in order to
be able to compare the knowledge the voting server gathers during
this process with the knowledge it gathers during the process
extended with the cast-as-intended mechanism.

\paragraph{Participants.}
The basic submission protocol is run among the following
participants: the voter $V$, the voting device $\VD$, and the
voting server $\VS$. In what follows, we implicitly assume that the channel from $V$
(via the voting devices) to the voting server $\VS$ is
authenticated without taking any assumption about how
authentication is carried out.\footnote{Since the exact method of authentication
is not relevant for the purposes of our cast-as-intended
protocol, we abstract away from authentication in our
presentation. In practice, the voter can use for example a
password to log in to $VS$.}

\paragraph{Cryptographic primitives.}

In the basic ballot submission protocol, an IND-CPA-secure public-key encryption scheme $\E = (\KeyGen, \Enc, \Dec)$ is employed.

\paragraph{Ballot submission (basic).}

We assume that $(pk, sk) \leftarrow \KeyGen$ was generated
correctly in the setup phase of the voting protocol and that each
party knows $pk$.\footnote{The secret key $sk$ is known only to the talliers of the election who use (their shares of) $sk$ to decrypt the ballots in the tallying phase. The exact method used to verifiably tally the ballots (via, e.g., homomorphic aggregation, or verifiable shuffling) is orthogonal to the cast-as-intended method proposed in this paper.} The program of the basic submission protocol works in the standard way:
\begin{enumerate}
 \item Voter $V$ enters plaintext vote $v$ to her voting device
     $\VD$.
 \item Voting device $\VD$ chooses randomness $r \xleftarrow{\$}
     R$, computes ciphertext $c \leftarrow \Enc(pk, v; r)$, and
        sends $c$ to voting server $VS$.
\end{enumerate}
We note that the basic protocol may include signing the ballot
with voter's private key if a public-key infrastructure (PKI)
among the voters is established.\footnote{Since this aspect is
independent of our cast-as-intended protocol, we do not assume
that voters sign their ballots in our presentation. We note that
our protocol also works with ballots signed by voters.}

\subsection{Cast-as-intended verifiable ballot submission}\label{sec:generic-protocol-extended}

We now describe how to extend the basic ballot submission
protocol described above for cast-as-intended verifiability.

\paragraph{Participants.}

In addition to the three participants of the basic ballot
submission phase (voter $V$, voting device $\VD$, voting server
$\VS$), the extended protocol also includes an audit device $\AD$.

\paragraph{Cryptographic primitives.}

The extended submission protocol employs the following cryptographic primitives:
\begin{enumerate}
 \item An IND-CPA-secure public-key encryption scheme $\E = (\KeyGen, \Enc,\allowbreak \Dec)$ that allows for re-randomization and special decryption:
 \begin{itemize}
  \item \emph{Re-randomization} guarantees the existence of a
      \emph{probabilistic poly\-nomial-time (ppt)} algorithm
         $\ReRand$ which takes as input a public key $pk$
         together with a ciphertext $c = \Enc(pk, m; r)$ and
         returns a ciphertext $c^*$ such that $c^* = \Enc(pk, m;
         r^*)$ for some (fresh) randomness $r^*$. We assume that
         $\ReRand$ is \emph{homomorphic} w.r.t.~randomness:
         $ 
           \Enc(pk, m; x + r) = \ReRand(pk, \Enc(pk, m; r); x).
         $
  \item \emph{Special decryption} guarantees the existence of a
      \emph{polynomial-time (pt)} algorithm $\Dec'$ which takes
         as input a public key $pk$, a ciphertext $c$, and a
         randomness $r$, and returns the plaintext
         $m$, if $c = \Enc(pk, m; r)$, or fails
         otherwise.\footnote{Special decryption
         is given for free if the message space is polynomially
         bounded: one can simply brute-force all the potential
         plaintext messages and encrypt each with the given
         randomness until this produces $c$.} 
  \end{itemize}
  Note that neither $\ReRand$ nor $\Dec'$ require knowledge of the secret key $sk$ associated to $pk$.
  \item A proof of correct re-encryption, i.e., an interactive
     zero-knowledge proof (ZKP) $\pi_{\ReRand}$ for the following relation:
  $
  (pk, c, c^*; x) \in \mathcal{R}_{\ReRand} \Leftrightarrow c^* = \ReRand(pk, c; x).
  $

  The joint input of the prover and the verifier is statement
  $(pk, c, c^*)$ and the secret input of the prover is
  witness $x$, i.e., the randomness used to re-randomize
  ciphertext $c$ into $c^*$.

\end{enumerate}

\paragraph{Ballot submission (extended).}
The program of the extended ballot submission works as follows (note that the first two steps are the ones of the basic ballot submission protocol):
\begin{enumerate}[label=(BS\arabic*),leftmargin=*]
 \item Voter $V$ enters plaintext vote $v$ to voting device $\VD$.
 \item 
     \label{step:cast}
     Voting device $\VD$ chooses randomness $r \xleftarrow{\$}
     R$, computes ciphertext $c \leftarrow \Enc(pk, v; r)$, and
    sends $c$ to voting server $\VS$.
 \item 
     \label{step:blinding-factor}
     Voting server $\VS$ chooses a blinding factor $x
     \xleftarrow{\$} R$ and sends $x$ to $\VD$.
 \item 
     \label{step:cast-final}
     Voting device $\VD$ computes blinded randomness $r^*
     \leftarrow x + r$ and returns $r^*$ to voter $V$ (in the
        practical implementations, $r^*$ can be displayed as a
        QR-code).
\end{enumerate}
From the voter's perspective, the outcome of the submission
protocol consists of the blinded randomness $r^*$, which is used for
individual verification purposes, as described next.

\paragraph{Cast-as-intended verification.}

The program of the voter's individual cast-as-intended
verification works as follows. It is executed, if the voter
chooses to audit his/her ballot. As for the ballot submission,
in what follows, we implicitly assume that the channel from $V$
(via the audit devices) to the voting server $VS$ is
authenticated.

\begin{enumerate}[label=(BA\arabic*),leftmargin=*]
    \item Voter $V$ enters $r^*$ to the audit device $AD$ (in
        practical implementations this is done by scanning a QR
        code produced by $\VD$), which contacts voting server $\VS$.
   \item \label{step:sd-initial}
     Voting server $\VS$ computes ciphertext $c^* \leftarrow
     \ReRand(pk, c; x)$ (i.e., original ciphertext $c$
      re-randomized with the blinding factor $x$) and sends the
      original ciphertext $c$ along with $c^*$ to the audit 
        device $AD$.
 \item Voting server $\VS$ and audit device $\AD$ run interactive
     zero-knowledge proof $\pi_{\ReRand}$, where $\VS$ is the
        prover and $\AD$ the verifier, with joint input $(pk, c,
        c^*)$ and voting server's secret input $x$ in order to
        prove/verify that $c^*$ is a re-randomization of $c$.
 \item If the verification algorithm in the step above 
      returned $1$, then $\AD$ decrypts the
        re-randomized ciphertext $c^*$ using blinded randomness
        $r^*$ to obtain $v^* \leftarrow \Dec'(pk, c^*, r^*)$ and
        returns $v^*$ to voter $V$. Otherwise, $\AD$ returns 0
        (indicating failure) to $V$.
    \item Voter $V$ returns 1 (accepts) if $AD$ returned $v^*$
        such that $v = v^*$ (where $v$ is the voter's intended choice).
        Otherwise, $V$ returns 0 (reject).
\end{enumerate}

%% file: security.tex
\section{Security}\label{sec:security}

Our cryptographic security analysis of the cast-as-intended protocol (as introduced in Sec.~\ref{sec:generic-protocol-extended}) consists of two parts. In the first part, we prove that this protocol is an interactive zero-knowledge proof (ZKP) protocol, run between voter $V$ and audit device $\AD$, jointly playing the role of the verifier on the one side, and the voting device $\VD$ and voting server $\VS$ jointly playing the role of the prover on the other side.
This fact establishes the \emph{cryptographic deniability} of our cast-as-intended method: the protocol transcript (the data gathered by the audit device) is useless as a receipt, because an indistinguishable transcript can be generated by any party, using the simulator algorithm (for an arbitrary election choice, independently of the actual voter's choice).

In the second part, we prove that the voting server $\VS$ does not learn more information about the voter's secret choice than what $\VS$ already learns in the basic ballot submission protocol.  Note that this statement is not directly covered by the zero-knowledge (simulation) property of the protocol, because $\VS$ is part of the prover.

In Sec.~\ref{sec:individual-verifiability}, we will explain how to extend the cast-as-intended protocol analyzed in this section so that it provides full individual verifiability.

\subsection{Zero-knowledge proof}

We that show our cast-as-intended protocol is a ZKP to prove that a given ballot contains a vote for a particular candidate. From the soundness of this ZKP, it follows that even if the voter's voting device $\VD$ and the voting server $\VS$ collude, then they are not able to convince the voter $V$ (who uses an honest audit device $\AD$) that her submitted ballot contains a vote for her favorite choice $v$ when it actually contains a different choice. Moreover, due to the zero-knowledge property, $\VD$ and $\VS$ prove that the submitted ballot contains a vote for the voter's favorite choice without revealing any information beyond this statement; in particular, the protocol does not leave any information which could undesirably serve as a receipt that could be used for vote buying.

Let $\Verify$ be the composition of the programs run by voter $V$
and her audit device $\AD$ after the basic ballot submission
protocol is completed, i.e., steps \ref{step:blinding-factor}--\ref{step:cast-final} in the extended ballot submission protocol followed by the cast-as-intended protocol; in
short: $\Verify = (V \| \AD)$. Analogously, let $\Prove$ be the
unification of the programs run by the voting device $\VD$ and the
voting server $VS$ after the basic ballot submission protocol is
completed; in short $\Prove = (\VD \| VS)$.

Observe that the resulting interactive protocol with joint input
$(pk, v, c)$ and prover's secret input $r$ can be re-written as
the following protocol:
\begin{enumerate}
 \item $\Prove$ chooses $x \xleftarrow{\$} R$, computes $r^*
     \leftarrow x + r$ and $c^* \leftarrow \ReRand(pk, c; x)$,
        and returns $(r^*, c^*)$.
 \item $\Prove$ and $\Verify$ run the interactive ZKP
     $\pi_{\ReRand}$ with joint input $(pk, c, c^*)$ and prover's
        secret input $x$.
 \item $\Verify$ returns 1 if and only if the execution of
     $\pi_{\ReRand}$ returned 1 and $v = \Dec'(pk, c^*, r^*)$
        holds true.
\end{enumerate}
We now state that this protocol is an interactive ZKP for proving that ciphertext $c$ encrypts vote $v$ (see Appendix~\ref{sec:zkp}, where we recall the definition of zero-knowledge proofs).

\begin{theorem}
 The interactive protocol $\pi_{\Enc} = (\Verify, \Prove)$ is a zero-knowledge proof for relation
 $(pk, v, c; r) \in \mathcal{R}_{\Enc} \Leftrightarrow c = \Enc(pk, v; r).$
\end{theorem}

In order to prove this theorem, we need to show that $\pi_{\Enc}$ satisfies correctness (i.e., if $\Verify$ and $\Prove$ are executed correctly for a true statement, then $\Verify$ returns 1), soundness (i.e., if $\Verify$ returns 1, then the statement is correct), and zero-knowledge (i.e., the verifier's view can be simulated without knowledge of the witness), each with at least overwhelming probability.

\begin{proof} 
 \emph{Correctness}: Let $x, x^*, c^*$ be defined as in $\Prove$. Because $(pk, c^*, c; x) \in \mathcal{R}_{\ReRand}$, the verifier returns 1 in an execution of $\pi_\ReRand$ with probability $p_c$, where $p_c$ is the correctness level of $\pi_\ReRand$. Furthermore, the verifier's second check is also positive because
 \begin{align*}
  c^* 
  = \ReRand(pk, \Enc(pk, v; r), x)
  = \Enc(pk, v; x + r)
  = \Enc(pk, v; r^*).
 \end{align*}
 Hence, $\Verify$ returns 1 in $\pi_{\Enc}$ with probability $p_c$ if both $\Verify$ and $\Prove$ are executed correctly; in short: $Pr[ \langle \Verify, \Prove(r) \rangle (pk, v, c) = 1 ] = p_c$.
 
 \emph{Soundness}: Assume that $\Verify$ returns 1. Then, due to
    the soundness of $\pi_{\ReRand}$, there exists with
    probability $p_s$ a unique plaintext $v^*$ such that we have $c^*
    \in \Enc(pk, v^*)$ and $c \in \Enc(pk, v^*)$, where $p_s$ is
    the soundness level of $\pi_{\ReRand}$. Furthermore, since
    $\Verify$ returns 1, by the property of special decryption
    $\Dec'$, we have $c^* \in \Enc(pk, v)$ and hence
    $v = v^*$. This means that
    $c \in \Enc(pk, v)$ with probability $p_s$.
 
 \emph{Zero-knowledge}: We can construct a simulator $\Sim$,
    which does not have access to the witness $r$ and which
    replaces $\Prove$ in the re-written protocol, as follows:
 \begin{enumerate}
 \item $\Sim$ chooses $r^* \xleftarrow{\$} R$, computes $c^* \leftarrow \Enc(pk, v; r^*)$, and returns $(r^*, c^*)$.
 \item $\Sim$ simulates the interactive ZKP $\pi_{\ReRand}$ without knowledge of $x$.
 \end{enumerate}
 Due to the ZK property of $\pi_{\ReRand}$, the verifier is not able to distinguish a real execution and a simulated one with probability $p_z$, where $p_z$ is the ZK level of $\pi_{\ReRand}$.
\end{proof}

\subsection{Simulatability towards voting server}

Recall that in the basic ballot submission protocol, the only
data that $\VS$ obtains from the voter is the voter's encrypted
choice $c = \Enc(pk, v; r)$. Due to the semantic security of the
public-key encryption scheme $\E$, the probability that $\VS$ can
derive any information about the voter's vote $v$ is negligible
(if $\VS$ is computationally bounded).

Now, in what follows, we show that the voting server $\VS$ does
not learn more information about the voter's vote in the
cast-as-intended protocol than what $\VS$ learns in the basic
ballot submission protocol. To this end, we compare the voting
server's view in both protocols and show that all additional
interaction between those participants that know/learn the voter's
vote (i.e., voter $V$ herself, her voting device $\VD$, and
her audit device $\AD$) on the one side and the voting server $\VS$
on the other side can be perfectly simulated without any
knowledge of the voter's vote $v$.

From the voting server's perspective, the basic ballot submission
protocol can be re-written as follows, where $\hat{V}$ is the
unification of the programs of $V$ and $\VD$:
\begin{enumerate}
 \item $\hat{V}$ chooses randomness $r \xleftarrow{\$} R$,
     computes ciphertext $c \leftarrow \Enc(pk, v; r)$, and sends
     $c$ to voting server $\VS$.
\end{enumerate}
From the voting server's perspective, the cast-as-intended
protocol (i.e., verifiable ballot submission followed by
cast-as-intended verification) can be re-written as follows,
where $\hat V_{ext}$ is the unification of the programs of $V$, $\VD$,
and $\AD$:
\begin{enumerate}
    \item $\hat V_{ext}$ chooses randomness $r \xleftarrow{\$} R$,
     computes ciphertext $c \leftarrow \Enc(pk, v; r)$, and sends
        $c$ to voting server $\VS$.
 \item Voting server $\VS$ chooses blinding factor $x
     \xleftarrow{\$} R$, computes ciphertext $c^* \leftarrow
        \ReRand(pk, c; x)$, and sends $(c^*,x)$ to voting device
        $\hat V_{ext}$.
 \item $\VS$ and $\hat V_{ext}$ run interactive ZKP $\pi_{\ReRand}$
     with joint input $(pk, c, c^*)$ and voting server's secret
        input $x$ in order to prove/verify that $c^*$ is a
        re-randomization of $c$.
\end{enumerate}
Due to the re-written presentations of the two protocols, it is
easy to see that from the voting server's perspective, the only
task carried out by $\hat{V}$ in the cast-as-intended protocol in
addition to $\hat V_{ext}$'s tasks in the ballot submission protocol
is executing the verification program of the interactive proof
$\pi_{\ReRand}$. Observe that the verification program of
$\pi_{\ReRand}$ can be executed by \emph{any} party which knows
$(pk, c, c^*)$; in particular no knowledge about the voter's vote
$v$ or randomization elements $r, r^*$ is required. We can
therefore perfectly simulate $\hat{V}_{ext}$'s additional program in
the cast-as-intended protocol. Using the standard (simulation)
argument that the voting server $\VS$ could run the simulation
algorithm (in our case: the verification program of
$\pi_{\ReRand}$) itself, we conclude that the voting server $\VS$
does not learn more information about the voter's vote in the
cast-as-intended protocol than what $\VS$ learns in the basic
ballot submission protocol.

\begin{remark}
    In the individually verifiable ballot submission protocol
    described above, the voting server $\VS$ does not learn
    whether the voter accepted or rejected a protocol run, i.e.,
    whether $\hat V_{ext}$ returned 0 or 1. Depending on the
    overall voting protocol specification, $\VS$ may however learn
    the  final output of $\hat V_{ext}$, for example, when the
    voting protocol requires that each voter submits a
    \emph{confirmation code} to the voting server after she
    completed her cast-as-intended verification successfully in
    order to publicly confirm that $V$ accepts the submitted
    ballot (see, e.g.,~\cite{GalindoGP15}).
 
    We note that even if $\VS$ learns the output of $\hat V_{ext}$,
    ballot privacy towards a possibly corrupted $\VS$ is still
    guaranteed in our cast-as-intended protocol. In order to
    prove this claim, we show that the probability of the event
    that the execution of $\pi_{\ReRand}$ returned 1 but $v \neq
    \Dec(pk, c^*, r + \tilde{x})$ holds true, where $(c^*,
    \tilde{x})$ is the output of $\VS$, is negligible. Let us
    consider the set of runs in which this event holds true. Due
    to the soundness of $\pi_{\ReRand}$, there exists $x \in
    \mathcal{R}$ such that $c^* = \ReRand(pk, c; x) = \Enc(pk, v;
    r+x)$. Now, if $v \neq \Dec'(pk, c^*, r + \tilde{x})$, then
    there exists $\tilde{v} \neq v$ such that $c^* = \Enc(pk,
    \tilde{v}; r+\tilde{x})$ holds true. Due to the correctness
    of the PKE scheme $\E$, it follows that $v = \Dec(sk, c^*) =
    \tilde{v}$, which is a contradiction to $v \neq \tilde{v}$.
 
    We can therefore conclude that the slightly extended
    cast-as-intended protocol can be simulated (with overwhelming
    probability) exactly as in the case above where $\VS$ does not
    learn the output of $\hat V_{ext}$ when we additionally
    specify that the simulator returns 1 to $\VS$ if and only if
    $\pi_{\ReRand}$ returns 1. Note that the simulator does not
    need to check whether $v = \Dec(pk, c^*, r^*)$ and hence does
    not need to know $v$.
\end{remark}

\section{Full Individual Verifiability
\label{sec:individual-verifiability}}

In the previous two sections, we presented the method for
cast-as-intended verifiability and analyzed the security
properties of this method. Cast-as-intended, which enables the
voter to audit his/her ballot and check that it contains the
intended choice, does not, however, fully cover the notion of
individual verifiability. What is missing is the guarantee that
the audited ballot takes part in the tally (sometimes called
\emph{tallied-as-recorded}).

In this section, we add the standard mechanism to achieve tallied-as-recorded verifiability: a public bulletin board and
signed receipts. We also state the higher level security
properties such a final system provides.

The content of this section can be seen as an example for how our
cast-as-intended mechanism can be embedded in a more complete
protocol to provide full individual verifiability.

As noted, we introduce an additional participant: the public
bulletin board. It is used to collect all the cast ballots, where
ballots are published together with unique voter identifiers.
We assume that the voters (or auditors) have access to this
public bulletin board (during and/or after the ballot cast
process) and can check that a given ballot is included there.

We also assume that the voting server has a (private) signing key
and that the corresponding (public) verification key is publicly
known.

The modifications to the protocol presented in
Section~\ref{sec:generic-protocol} are straightforward.
The changes in the ballot submission protocol are as follows.
\begin{itemize}
\item
    The encrypted ballot $c$ submitted in Step
    \ref{step:cast} is published by the voting
    server on the public bulletin board together with a unique
    voter's identifier.
\item
    In step \ref{step:blinding-factor}, the voting server $\VS$ 
    additionally sends to the voting device $\VD$ \emph{a signed
    ballot cast confirmation $s$}, that is a signature on the
    cast ballot $c$. The signature $s$ is then checked by
    the voting device $\VD$ and $s$ is given to the voter in Step
    \ref{step:cast-final}.
\end{itemize}
We also consider the following changes in the ballot audit
process:
\begin{itemize}
\item
    The voting server $\VS$, in Step \ref{step:sd-initial}, sends
    additionally to the audit device $\AD$ the ballot cast
    confirmation $s$, as in the step above. The audit device
    checks that $s$ contains a valid signature of the voting
    server on $c$.
\item
    In the final step of the ballot audit process, the voter is
    given the signed ballot cast confirmation.
\end{itemize}
Note that the ballot cast confirmation is provided to the
voter twice: once by the voting device and then by the audit
device. It is expected that these confirmations are exactly
the same (which is the case when both devices are honest).

With such receipt, the voter, having executed the ballot audit
process, has the following guarantees which directly follow from
the results of Section \ref{sec:security}.

\begin{theorem}[informal]
    Assume that at least one of the voter devices (the voting
    device or the audit device) is honest.
    If the voter successfully carried out the ballot cast process
    and the ballot audit process, then the voter is in the possession of ballot
    confirmation which (1) is correctly signed by the voting server, and (2)
    refers to an encrypted ballot containing the voter's
    intended choice (as shown to the voter and confirmed in
    the ballot audit process).
    
    At the same time, the second device (even if it behaves
    dishonestly) is not able to produce a convincing evidence
    for a third party about the voter's choice.
\end{theorem}

With this result, given that one of the devices is honest, the
voter can check that their ballot, containing their intended
choice, is included in the public bulletin board (and if not,
given the valid signature, the voter can demonstrate that the
voting server misbehaved) and by this also included in the final
tally (where the correctness of the tallying process is given due
to the universal verifiability).

Note that to strengthen this result, the voter can even carry out
the ballot audit process using more than one device. With this,
even if only one of these devices was honest, it would be enough
to guarantee cast-as-intended.

%% file: instantiations.tex
\section{Instantiations}\label{sec:instantiations}

We demonstrate that our cast-as-intended protocol can be instantiated with common cryptographic primitives. Our protocol can therefore be used to extend important e-voting protocols for cast-as-intended verification.

\subsection{ElGamal-based e-voting schemes}\label{sec:elgamal}

    A natural instantiation of our method is the one based on the
    standard ElGamal group of order $q$ with a generator $g$. 
    In this setting, the public key is of the form $h = g^{sk}$,
    where $sk \in Z_q = \{0,\dots,q-1\}$. Given a plaintext message
    message $m \in Z_q$, the encryption of $m$ with randomness
    $r$ is $c = (g^r, m\cdot h^{r})$.

    \emph{Special decryption}:
    For a ciphertext of the form $c = (u,w)$ encrypted using
    randomness $r$ (which means that $u = g^r$ and $w = m\cdot
    h^{r}$), the randomness $r$ allows one to easily extract the
    plaintext message by (checking that $u$ is in fact $g^r$ and)
    computing $w \cdot h^{-r}$.

    \emph{Re-randomisation} of a ciphertext $c = (u,w)$ is of the
    form $c' = (u', w')$ where $u' = u \cdot g^x$ and $w' = w
    \cdot h^x$. In order to prove that $c'$ is a re-randomisation
    of $c$, one can use the well-known sigma-protocol for
    equality of discrete logarithms, that is the proof of
    knowledge of $x$ such that $X = \frac{u'}{u} = g^x$ and $Y =
    \frac{w'}{w} = h^x$~\cite{chaum1992wallet},
    and transform it into an interactive zero-knowledge protocol using,
    for instance, the technique from~\cite{hazay2010sigma, lindell2018zero}.
    See Appendix~\ref{sec:zkp-eq}, where, for the sake of completeness,
    a specific instantiation is provided.


    We note that the \emph{computational cost} of this method is
    low and the protocol can, therefore be easily handled even by
    low-end general purpose devices: There is essentially no
    extra cost on the voting device (no additional modular
    exponentiations). On the server (prover) side, the ballot
    audit process requires 6 modular exponentiations (2 for
    re-randomisation and 4 for the ZKP).  The audit device
    (verifier) needs 8 modular exponentiations: 6 for the ZKP and
    2 for special decryption To put this number in a perspective,
    it is comparable to the cost of ballot preparation in a
    typical ElGamal-based voting system which, in the simplest
    case, requires 3 modular exponentiations. For an
    implementation using elliptic-curve-based ElGamal group, on
    an Android phone with a relatively modern CPU (Qualcomm®
    Snapdragon 865 CPU) the ballot audit process takes only
    roughly 0.08 seconds, for a simple ballot which can be encoded as
    one group element, and it scales linearly with ballot length.

\subsection{Commitment-based e-voting schemes}


E-voting protocols, in which the voters' choices are publicly
``encrypted'' not as ciphertexts but as commitments, can offer
several advantages. For
example,~\cite{DBLP:conf/eurosp/BoyenHM21,DBLP:conf/ccs/PinoLNS17}
provide practical post-quantum security,
and~\cite{DBLP:conf/esorics/CuvelierPP13,DBLP:conf/eurocrypt/CramerFSY96}
guarantee everlasting privacy. In what follows, we will
explain that our generic cast-as-intended protocol
(Sec.~\ref{sec:generic-protocol-extended}) can easily be adapted
to this family of voting protocols.

Recall that in the basic ballot submission protocol described in
Sec.~\ref{subsec:generic-protocol-basic}, each voter's choice $v$
is encrypted by the voting device $\VD$ as a ciphertext $c
\leftarrow \Enc(pk, v; r)$. Now, in the basic ballot submission
phase of a voting protocol where voters commit to their votes,
the voting device $\VD$ ``encrypts'' the voter's choice $v$ as
a commitment $c \leftarrow \Com(v; r)$, where $r$ are the
random coins (as before), and sends $c$ to the voting server
$\VS$.\footnote{In fact, the voting device also encrypts the
opening values under the talliers' public key $pk$ and sends the
resulting ciphertext to $\VS$ or privately to the talliers,
respectively. Since such details are not relevant for the
purposes of this paper, we will omit them in what follows.} 

Analogously to our cast-as-intended verifiable protocol in
Sec.~\ref{sec:generic-protocol-extended}, we can extend the basic
ballot submission phase of commitment-based voting protocols to
provide cast-as-intended verification, as described next.
Regarding the cryptographic primitives, we also need to assume
that the commitment scheme (just like the encryption scheme in
Sec.~\ref{sec:generic-protocol-extended}) allows for
re-randomization, that we can use an interactive
zero-knowledge proof for proving correct re-randomization of
commitments, and that a message in a commitment can be obtained
using the randomness of the commitment.

All of the commitment-based voting protocols mentioned
above (i.e.,
~\cite{DBLP:conf/eurosp/BoyenHM21,DBLP:conf/ccs/PinoLNS17,DBLP:conf/esorics/CuvelierPP13,DBLP:conf/eurocrypt/CramerFSY96})
provide the necessary cryptographic features
(re-randomization of commitments, ZKP of correct
re-randomization, and reconstruction of committed messages via randomness); moreover, these protocols and the primitives they employ were proven practically efficient in the respective publications. Therefore, all of the aforementioned important commitment-based protocols can easily be extended with our
cast-as-intended protocol to provide individual verifiability in real-world elections.

%% file: zkp.tex
\section{Zero-Knowledge Proofs}\label{sec:zkp}

We recall the definition of interactive zero-knowledge proof (ZKP) systems. Since each interactive ZKP is an interactive proof system (PS), we define this primitive first.

On a high level, \emph{completeness} of a PS ensures that, if both the prover $P$ and the verifier $V$ are honest, then for each true statement proven by the prover, the verifier should accept the proof (with overwhelming probability). \emph{Soundness} of a PS guarantees that a possibly dishonest prover $B$ is not able to convince a verifier $V$ that a false statement was true.

In our setting, we consider languages in NP, and so the set of `true'
statements of a given PS is defined as the set
$L_\mathcal{R} = \{ x \mid (x,w) \in \mathcal{R} \text{ for some $w$}\}$,
for some relation $\mathcal{R}$ decidable in polynomial time.
For $(x,w) \in \mathcal{R}$, $x$ is called a statement and $w$ is called a witness. By convention, we will represent such pairs using semicolons, as $(x; w)$.

\begin{definition}[Interactive Proof System]
 At tuple $(P, V)$ is an \emph{interactive proof system} for a relation $\mathcal{R}$ if $(P, V)$ is a pair of connected interactive Turing machines (ITMs), $V$ has polynomial runtime, and the following two conditions hold true:
 \begin{itemize}
  \item \emph{Completeness:} For all $(x;w) \in \mathcal{R}$, the probability $Pr(P(x,w) \mid V(x) = 1)$ is overwhelming. That is, on common input $x$, if the honest prover gets as private input $w$ such that $(x; w) \in \mathcal{R}$, then the honest verifier accepts with overwhelming probability.
  \item \emph{Soundness:} For all $x \notin L_\mathcal{R}$ and all ITMs $P^*$ connected to $V$, $Pr(P^*(x) \mid V(x) = 1)$ is negligible. That is, the probability that the honest verifier $V$ accepts a false statement, when interacting with any (dishonest) prover $P^*$ is negligible.

    If we restrict $P^*$ to be a probabilistic polynomial-time (PPT) Turing machine, then the definition yields \emph{computational soundness}.

 \end{itemize}
\end{definition}

Now, an interactive ZKP is an interactive PS in which one more property is provided: essentially, the verifier $V$ does not learn more information about the prover's secret from the conversation with $P$ than the fact that the statement is true. Formally, this concept is defined via the notion of a simulator.

\begin{definition}[Zero-Knowledge Proof System]
 Let $(P,V)$ be an interactive PS for $\mathcal{R}$. Then, $P$ is called \emph{(computational) zero-knowledge (ZK)}, if for all PPT ITMs $V^*$ connected to $P$, there exists an expected probabilistic polynomial-time (expected PPT) algorithm $S$, called a \emph{simulator}, such that, for all $(x;w) \in \mathcal{R}$, the distribution of the communication transcripts of executions of $P(x,w)$ with $V^*(x)$ is (computationally) indistinguishable from the output of $S$ on $x$.
\end{definition}

%% file: zkp-eqlog.tex
\section{ZKP of Equality of Discrete Logarithms} \label{sec:zkp-eq}

In this section, we provide a concrete instantiation of the zero-knowledge
protocol for equality of discrete logarithms.
As in Section~\ref{sec:elgamal}, we take the standard ElGamal
group of order $q$ with a generator $g$.

The standard sigma protocol which proves the knowledge of such an $x$
that $X = g^x$ and $Y = h^x$ works as follows~\cite{chaum1992wallet}.
The prover (who knows the shared discrete logarithm $x$)
samples random $a \in Z_q$ and sends
$A = g^a$ and $B = h^a$
to the verifier who replies with a random challenge
$e \in Z_q$.
The prover replies with the value
$ z = a + ex$ and the verifier accepts the proof if
$A = \frac{g^z}{X^e}$ and $B = \frac{h^z}{Y^e}$.

As this sigma protocol is only \emph{honest-verifier zero knowledge}
and is not known to provide the zero-knowledge property in
the general case, it cannot be directly used in our cast-as-intended
protocol. We can, however, apply the technique from~\cite{hazay2010sigma} to
obtain an efficient interactive zero-knowledge protocol.
The high-level idea is that the verifier, first, commits to the
challenge $e$, using a perfectly hiding commitment scheme, before the original
sigma protocol is carried out.

For concreteness, we instantiate this technique with the Pedersen commitments,
obtaining the following protocol:

\begin{enumerate}
\item
    In the initial step, the prover samples random $\tau \in Z_q$ and sends
    \[ k = g^\tau \]
    to the verifier (the value $k$ will serve as the commitment key).
\item
    The verifier samples random $e, r \in Z_q$ and sends
    \[ c = g^r k^e \]
    to the prover ($c$ is a commitment to $e$ with the randomization factor $r$
    and $e$ is intended to be used as the challenge in Step~\ref{zkpeq-challenge}).
\item
    The prover, as in the original sigma protocol above,
    samples random $a \in Z_q$ and sends
    \[A = g^a \quad\text{and}\quad B = h^a.\]
\item \label{zkpeq-challenge}
    The verifier decommits to the challenge $e$ by sending
    \[ e,\, r \]
\item
    If the decommitment is not correct (that is if $c \neq g^r k^e$),
    the prover aborts. Otherwise, it replies with
    \[
        z = a + ex
    \]
\item
    As above, the verifier accepts the proof if
    \[
        A = \frac{g^z}{X^e}
        \quad\text{and}\quad
        B = \frac{h^z}{Y^e}.
    \]
\end{enumerate}

\paragraph{Security.}

By the results of~\cite{hazay2010sigma, lindell2018zero}, this protocol
is a zero-knowledge protocol. Because in our instantiation we use Pedersen
commitments, the result (more specifically, the zero-knowledge property) holds
under the assumption that the DLOG problem is hard.

\paragraph{Computational cost.}
In the above protocol, the prover needs to compute 4
modular exponentiations (if the commitment key $k$ is reused), while
the verifier computes 6 modular exponentiations (the
remaining operations have a negligible effect). These numbers, especially
for elliptic cure groups, yield a very efficient protocol.
It means that, on one hand side, the protocol does not add much extra
computational cost on the election server side (prover) and,
on the other hand side, the ballot audit procedure (the
verification) can be easily handled even by low-end general
purpose devices (even when a ballot it too big to be
encoded as one element of $Z_q$ and, therefore, the
zero-knowledge protocol needs to be executed some some number
of times).